\documentclass{biometrika}

\usepackage{amsmath, amssymb}

\usepackage{newtxtext}
\usepackage[subscriptcorrection]{newtxmath}

\graphicspath{{./art/}}

\usepackage[plain,noend]{algorithm2e}

\makeatletter
\renewcommand{\algocf@captiontext}[2]{#1\algocf@typo. \AlCapFnt{}#2} 
\def\@algocf@capt@plain{top}
\renewcommand{\algocf@makecaption}[2]{%
  \addtolength{\hsize}{\algomargin}%
  \sbox\@tempboxa{\algocf@captiontext{#1}{#2}}%
  \ifdim\wd\@tempboxa >\hsize
    \hskip .5\algomargin%
    \parbox[t]{\hsize}{\algocf@captiontext{#1}{#2}}
  \else%
    \global\@minipagefalse%
    \hbox to\hsize{\box\@tempboxa}
  \fi%
  \addtolength{\hsize}{-\algomargin}%
}
\makeatother

\def\est{\hat{\beta}}
\def\scP{\mathcal{P}}
\def\mpm{\textsc{mpm}}
\def\apm{\textsc{apm}}
\def\bd{\textsc{bd}}
\def\bma{\textsc{bma}}
\def\fd{\textsc{fd}}
\def\is{\textsc{is}}
\def\fdp{\textsc{fdp}}
\def\tpp{\textsc{tpp}}
\def\mse{\textsc{mse}}
\def\top{{ \mathrm{\scriptscriptstyle T} }}

\usepackage{bbm}
\usepackage{hyperref}

\begin{document}

\markboth{A. Li, S. T. Tokdar \and J. Xu}{Sparse estimation by Bayesian decision theory}

\title{A Bayesian decision-theoretic approach to sparse estimation}

\author{Aihua Li}
\affil{Department of Statistical Science, Duke University,\\ 214 Old Chemistry, Durham, North Carolina 27708, U.S.A.
\email{aihua.li@duke.edu}}

\author{Surya T. Tokdar}
\affil{Department of Statistical Science, Duke University,\\ 214 Old Chemistry, Durham, North Carolina 27708, U.S.A.\email{surya.tokdar@duke.edu}}

\author{\and Jason  Xu}
\affil{Department of Biostatistics, University of California Los Angeles,\\ 650 Charles E. Young Dr. S., Los Angeles, California 90095, U.S.A.
\email{jqxu@g.ucla.edu}}

\maketitle

\begin{abstract}
We extend the work of \cite{hahn_decoupling_2015} and develop a doubly-regularized sparse regression estimator by synthesizing Bayesian regularization with penalized least squares within a decision-theoretic framework. In contrast to existing Bayesian decision-theoretic formulation chiefly reliant upon the symmetric 0-1 loss, the new method -- which we call Bayesian Decoupling -- employs a family of penalized loss functions indexed by a sparsity-tuning parameter. We propose a class of reweighted $l_1$ penalties, with two specific instances that achieve simultaneous bias reduction and convexity. The design of the penalties incorporates considerations of signal sizes, as enabled by the Bayesian paradigm. The tuning parameter is selected using a posterior benchmarking criterion, which quantifies the drop in predictive power relative to the posterior mean which is the optimal Bayes estimator under the squared error loss. Additionally, in contrast to the widely used median probability model technique which selects variables by thresholding posterior inclusion probabilities at the fixed threshold of $1/2$, Bayesian Decoupling enables the use of a data-driven threshold which automatically adapts to estimated signal sizes and offers far better performance in high-dimensional settings with highly correlated predictors. Our numerical results in such settings show that certain combinations of priors and loss functions significantly improve the solution path compared to existing methods, prioritizing true signals early along the path before false signals are selected. Consequently, Bayesian Decoupling produces estimates with better prediction and selection performance. Finally, a real data application illustrates the practical advantages of our approaches which select sparser models with larger coefficient estimates. 
\end{abstract}


\section{Introduction}

The statistical discourse on sparse linear regression is dominated by two main methodological paradigms. One paradigm relies on least squares estimation regularized by sparsity-encouraging penalties. The second builds probabilistic expressions of sparsity through a Bayesian framework. This paper investigates a decision-theoretic synthesis of these two paradigms with surprising benefits. 

Our starting point is the two-stage estimation method of \citet{hahn_decoupling_2015}, where a Bayesian model is used in the first stage to arrive at a posterior distribution on the regression coefficient vector. In the second stage, a sparse Bayes estimate is extracted as the minimizer of  posterior risk under a loss function which combines squared prediction error with an $l_1$ penalty. Operationally, the final estimate is simply an $l_1$-penalized least squares estimate where the lasso method \citep{tibshirani_regression_1996} is applied after replacing the original response data with their smoothed estimates produced by the first-stage Bayesian analysis. \citet{hahn_decoupling_2015} motivated this two-stage method as a means of  ``decoupling shrinkage and selection''. They considered the use of a shrinkage prior in the first stage for computational ease but burdened with the liability that no expression of sparsity is offered by the corresponding posterior. This necessitated a second stage where a loss function is brought into the picture to produce a sparse Bayes estimate via posterior risk minimization. 

Like the lasso and other regularized least squares methods, the second stage of the above approach produces a solution path as a function of a tuning parameter which controls the relative weight of the sparsity penalty vis-a-vis prediction error. But an important point of departure arises in selecting this parameter. Unlike regularized least squares, no cross-validation is required. The first-stage posterior distribution offers a ready quantification of the trade-off between prediction accuracy and sparsity gain, which is exploited to make a data-adaptive choice. This act of benchmarking against the posterior gives the method a strong Bayesian flavor and makes it computationally more tractable than many regularized least squares methods.

The scope of this two-stage method is more profound than its appearance as a computational hack for manufacturing sparsity within Bayesian shrinkage analyses. Indeed, shrinkage priors are no longer a computational necessity for high-dimensional Bayesian regression. Recent advances in computational techniques have made it possible to address very high-dimensional problems with spike-and-slab selection priors which directly encode sparsity in the analysis \citep[e.g.,][]{zanella_scalable_2019,griffin_2020}. However, while the spike-and-slab posterior quantifies uncertainty over variable selection, it may not facilitate sparse estimation. Indeed, the posterior mean of the coefficient vector -- the Bayes estimate minimizing posterior expected squared error loss, which is the foundation of many areas of Bayesian analysis including variable selection \citep[e.g.,][]{ishwaran_spike_2005} -- is entirely non-sparse. Its failure to produce exact zeros in the coefficient estimate makes it suboptimal when the true coefficient vector is sparse \citep{Johnstone2004}. This result is not surprising at all; Bayesian model fitting is separate from Bayesian estimation. A formal treatment of the latter requires minimizing the posterior risk under a suitably chosen loss function. Clearly, the characteristics of the estimates can depend as much on the chosen loss function as they do on the chosen prior.

This paper exploits this decision-theoretic interpretation of the two-stage method of \citet{hahn_decoupling_2015} and extends it in two major directions. First, we experiment with a general class of reweighted $l_1$ penalties as loss functions, where the reweighting formulation can be customized to the prior, with important structural differences arising between shrinkage and selection priors. We demonstrate that certain specific combinations of the prior and the loss yield high-quality sparse estimates while others -- including the estimate proposed by \citet{hahn_decoupling_2015} -- do not. Remarkably, the computational cost of these more intricate combinations remains on par with that of the original two-stage method. While both \citet{hahn_decoupling_2015} and our method generate solution paths in a sparsity-tuning parameter, the optimal value of the tuning parameter in our approach is determined by a posterior benchmarking criterion, which has easier interpretation and broader applicability than the criterion used by \citet{hahn_decoupling_2015}.

Second, we utilize the two-stage framework to augment prevailing decision-theoretic approaches to Bayesian sparse learning reliant on model selection. A common practice along this line is to summarize the spike-and-slab posterior into marginal posterior inclusion probability (\textsc{pip}) of each predictor variable, and then threshold the \textsc{pip}s to select a ``model'', i.e., a sparse subset of important variables. Setting the threshold at $1/2$ yields the median probability model (\mpm), which enjoys optimality properties under rather strong assumptions \citep{barbieri_optimal_2004}. On the other hand, the \mpm\ is known to perform poorly when the predictors are highly correlated, being prone to selecting an empty subset because the marginal \textsc{pip}s can all be low due to competition (see our numerical illustration in Section \ref{sec:apm}). An adaptive choice of the threshold would be useful here, but currently no consensus has been reached in the literature on how to choose one. We show that an excellent data-dependent choice can be obtained by embedding this model selection exercise within the two-stage decision-theoretic framework. Doing so produces a solution path in the threshold parameter in the second stage, followed by data-adaptive selection of the threshold by the use of suitable posterior benchmarking. 

Though motivated by Bayesian considerations, this two-stage method is a compelling sparse estimation technique in its own right. In particular, for a well-chosen combination of prior and loss, it offers a substantive improvement of the solution path compared to popular regularization methods. Indeed, lasso solution paths are known to contain a large fraction of false discoveries early on \citep{su_false_2017} -- a problem that even the most judicious choice of the penalty parameter cannot solve completely. Improvement of the solution paths requires adapting the scale of the penalty potentially heterogeneously across predictors, which is carried out typically with either some crude estimates of the coefficient vector \citep[e.g.,][]{candes_enhancing_2008, zou_adaptive_2006} or more complex penalty functions which require tuning potentially as many penalty parameters as the number of predictors \citep[e.g.,][]{bogdan_slopeadaptive_2015}. In contrast, under the two-stage method, truly important predictors tend to appear early on the path while potentially spurious associations are pushed to the back. This is largely due to the customization of the loss function made possible under the two-stage method: the expected loss under the first-stage posterior automatically creates a variable-adapted heterogeneous penalization for the second stage. Through this lens, the two-stage method can also be viewed as a doubly-regularized sparse estimation method, which appears to significantly improve performance over peer methods.

The remainder of the paper is organized as follows. Section \ref{sec:sparse_est} defines the two-stage estimation method augmented using an extended class of reweighted $l_1$ penalties. Two specific prior-loss combinations are introduced: one that combines the spike-and-slab prior with false discovery reweighted $l_1$ penalty, and another that combines the horseshoe shrinkage prior with an inverse signal strength $l_1$ penalty. Section \ref{sec:property} presents a theoretical analysis of their statistical properties in the orthogonal design setting. Computational methods for both stages are also discussed,  along with a new criterion function for choosing the tuning parameter introduced in Section \ref{sec:post_benchmarking}. Section \ref{sec:apm} introduces an adaptive thresholding approach by framing Bayesian model selection by \mpm\ within the two-stage formulation. Section \ref{sec:result1} presents results from a numerical study in which we assessed prediction and selection performance of the proposed methods relative to existing sparse learning alternatives. Section \ref{sec:result2} discusses the double-regularization property of the two-stage method and its impact on improving solution paths. Section \ref{sec:eQTL} presents a real-data case study with an application to eQTL mapping. Section \ref{sec:discussion} concludes with closing remarks.

\section{Sparse estimation by Bayesian decision theory}\label{sec:sparse_est}

\subsection{Bayes estimate under parametric family of loss}\label{sec:bayes_est}

We aim to derive a Bayes estimate of a $p$-dimensional coefficient vector $\beta=[\beta_1, \ldots, \beta_p]^\top$ in the linear regression model $y\sim N(X\beta,\sigma^2 I_n)$, where $y=[y_1, \ldots, y_n]^\top$ is an $n$-dimensional response vector, $X$ is a fixed $n\times p$ design matrix, $I_n$ is an $n\times n$ identity matrix, and $\sigma^2>0$ is unknown. Obtaining a Bayes estimate requires two additional theoretical components: a prior distribution $\Pi$ on the model parameters $(\beta, \sigma^2)$ and a loss function $L(\beta, b)$ that offers a quantitative scoring of the error incurred in estimating $\beta$ by $b \in \mathbb{R}^p$. The Bayes estimate $\est$ is the minimizer of the posterior risk $E\{L(\beta,b) \mid y\}$. A more general form of a loss function $L(y^*,\beta,b)$ could be considered to take into account predictions of a future independent observation $y^* \sim N(X^*\beta,\sigma^2I_n)$ at a known future design matrix $X^*$, with the Bayes estimate now defined as the minimizer of $E\{L(y^*,\beta,b) \mid y\}$.

Following \cite{hahn_decoupling_2015}, we further expand this framework to accommodate a family of loss functions $\{L_\lambda(y^*,\beta,b):\lambda \ge 0\}$ of the form 
\begin{equation}\label{loss-bd}
L_\lambda(y^*,\beta,b)=\|y^*-X^*b\|_2^2+\lambda\mathcal{P}(\beta,b),\quad \lambda \ge 0,
\end{equation}
which combine the $l_2$ prediction error loss $\|y^*-X^*b\|_2^2$ with a penalty term $\mathcal{P}(\beta,b)$ that becomes large when $b$ is dense and small when it is sparse. Consequently, optimizing over \eqref{loss-bd} requires the solution to compromise between prediction accuracy and sparsity gain. The formulation \eqref{loss-bd} is more general than the one considered by \cite{hahn_decoupling_2015} in that we allow the penalty term to depend on the unknown parameter $\beta$ for reasons that will become apparent with the specific examples of $\mathcal{P}(\beta,b)$ presented in Section \ref{sec:penalty}. 

Under the above loss specification, the posterior risk can be calculated by using the standard bias-variance decomposition:
$$
\begin{aligned}
  E\{L_\lambda(y^*,\beta,b)\mid y\}&=E\{\|X^*\beta-X^*b\|_2^2\mid y\}+E\{n\sigma^2\mid y\}+\lambda E\{\mathcal{P}(\beta,b)\mid y\}\\
  &=\|X^*\bar{\beta}-X^*b\|_2^2+\lambda E\{\mathcal{P}(\beta,b)\mid y\}+C.
\end{aligned}
$$
where $C$ represents the terms that do not depend on $b$ and $\bar{\beta}=E\{\beta\mid y\}$ is the posterior mean of $\beta$. For notational simplicity, the mathematical and numerical analyses in the rest of this paper take $X^*=X$, although we note that this assumption is not necessary for applying the proposed approach in practice. Denote the corresponding Bayes estimate as
\begin{equation}\label{est-bd}
  \est_\lambda\in \mathop{\arg\min}_{b\in\mathbb{R}^p}\;\|X\bar{\beta}-Xb\|_2^2+\lambda E\{\mathcal{P}(\beta,b) \mid y\}.
\end{equation}
Following the terminology in \cite{hahn_decoupling_2015}, we call \eqref{est-bd} \textit{Bayesian Decoupling (\bd)} estimate. In this paper, the implementation of \bd\ is restricted to convex $l_1$-type penalties, ensuring that the minimization problem in \eqref{est-bd} has a unique solution and is computationally tractable.

Expanding the logic of \cite{hahn_decoupling_2015}, we interpret Bayesian Decoupling as disentangling estimation from modeling. While the purpose of modeling is to restrict and bias possible explanations of the data, estimation should be treated as an independent exercise which is potentially agnostic to model or prior specification, and whose objectives are articulated through the design of the loss function. Two primary advantages of doing so are (1) the choice of the prior does not need to be constrained by the estimation objective, and (2) multiple estimation tasks can be performed simultaneously under a single posterior by using different loss functions. For example, under a spike-and-slab prior, one can report the posterior mean (i.e., Bayesian model averaging, \bma) estimate under squared error loss for optimal prediction, while also reporting the sparse estimate under loss \eqref{loss-bd} to highlight the most relevant predictors. 

A key feature of \bd\ is the use of a family of loss functions indexed by a sparsity-tuning parameter $\lambda \ge 0$. By varying the value of $\lambda$, this framework produces a sequence of sparse estimates, which we refer to as a {\it solution path}. A smaller value ($\lambda\to 0$) favors denser estimates and a larger value ($\lambda\to +\infty$) favors the null estimate. There are many criteria one could define to choose a specific estimate from the path. We find the following approach appealing: pick the sparsest estimate for which the loss in predictive accuracy remains within a prespecified tolerance level, where the loss in predictive accuracy is measured against the \bma\ estimate obtained under $\lambda = 0$. Section \ref{sec:post_benchmarking} presents a formal definition and algorithm which, we believe, simplifies the approach introduced in \cite{hahn_decoupling_2015}.

Although \bd\ arises from a Bayesian framework of sparse estimation, it is useful to interpret it as a general estimation method. \bd\ can be seen as a novel \textit{doubly-regularized estimator}, integrating Bayesian model averaging in the first stage with sparse penalization in the second stage. A key distinction from routine regularized estimation is that by utilizing a posterior distribution in the first stage, \bd\ has access to a useful probabilistic characterization of the unknown parameter value, which can be advantageous in the second stage in multiple ways. In the next few sections, we explore three such advantages of \bd: its ability to customize the loss function and thereby substantially improve performance (Section \ref{sec:penalty}), its ability to offer a data-driven choice of the tuning parameter $\lambda$ without cross-validation (Section \ref{sec:post_benchmarking}), and its ability to denoise and amplify important signals to produce higher-quality solution paths (Section \ref{sec:result2}).

\subsection{A class of reweighted $l_1$ penalties}\label{sec:penalty}

The design of $\mathcal{P}(\beta,b)$ should be such that sparsity can be induced in solving the minimization problem \eqref{est-bd}. A simple choice adopted by \cite{hahn_decoupling_2015} is the $l_1$ penalty, $\mathcal{P}(\beta,b)=\|b\|_1$, motivated by the lasso \citep{tibshirani_regression_1996}. It is a convex relaxation of the $l_0$ penalty, $\scP(\beta, b)=\|b\|_0$, and leads to the convex optimization problem
\begin{equation}\label{est-l1}
\est^{l_1}_\lambda=\mathop{\arg\min}_{b\in\mathbb{R}^p}\;\|X\bar{\beta}-Xb\|_2^2+\lambda\|b\|_1,
\end{equation}
offering benefits such as computational efficiency and continuous shrinkage \citep{tibshirani_regression_1996, fan_li_2001}. The estimate $\est^{l_1}_\lambda$ resembles the lasso estimate, except that the squared error term replaces the raw response vector $y$ with the \bma\ estimate $X\bar{\beta}$. 

It is well known that the use of $l_1$ penalty in the lasso introduces nonnegligible estimation bias, especially for large signals, due to uniform shrinkage applied across all coefficient estimates. This bias often forces the lasso estimate to include irrelevant variables with small coefficient estimates and thus an excessive number of false discoveries (e.g., see \citealp{Song_Liang_rLasso} and \citealp{fan_li_2001}). The same issue arises when applying the $l_1$ penalty within the \bd\ framework. 

One popular solution to the above problem is to let the penalty vary heterogeneously across the underlying true signals in a way such that larger signals receive less penalization. For example, \cite{candes_enhancing_2008} and \cite{zou_adaptive_2006} define a reweighted $l_1$ minimization with penalty $\sum_{j=1}^p w_j|b_j|$, where the weight $w_j$ depends inversely to the signal strength. \cite{fan_li_2001} proposes the smoothly clipped absolute deviation (\textsc{scad}) penalty, which is bounded by a constant for large signals to provide nearly unbiased estimates. \cite{Song_Liang_rLasso} defines the reciprocal Lasso (\textsc{rlasso}) penalty, $\sum_{j=1}^p|b_j|^{-1}\mathbbm{1}(b_j\neq 0)$, which vanishes to zero as the estimate $b_j$ increases in magnitude. However, these approaches come with various drawbacks. In reweighted $l_1$ minimization, ambiguity remains in the choice of the weight sequence, which importantly conveys a prior ranking of variable importance; current strategies take on iterative algorithms \citep{candes_enhancing_2008} or ordinary least squares surrogate \citep{zou_adaptive_2006}. The \textsc{scad} and \textsc{rlasso} penalties sacrifice convexity and thus confront greater computational challenges. In this section and Section \ref{sec:property}, we report the somewhat surprising finding that simultaneous bias reduction and convexity can be achieved by \bd\ with fairly neat penalty designs. 

We define a class of reweighted $l_1$ penalties as
\begin{equation}\label{penalty-reweighted-l1}
  \mathcal{P}(\beta,b)=\sum_{j=1}^p w_j(\beta)|b_j|,
\end{equation}
where $\{w_j(\beta)\}\subset\mathbb{R}$ is a sequence of nonnegative weights depending on the unknown true parameter $\beta$. Evaluating the posterior expectation of the \bd\ loss \eqref{loss-bd} with the penalty \eqref{penalty-reweighted-l1} leads to the \bd\ estimate 
\begin{equation}\label{est-reweighted-l1}
  \est^{\text{Reweighted}}_\lambda=\mathop{\arg\min}_{b\in\mathbb{R}^p}\;\|X\bar{\beta}-Xb\|_2^2+\lambda\sum_{j=1}^p\hat{w}_j|b_j|,
\end{equation}
where $\hat{w}_j=E\{w_j(\beta)\mid y\}\ (j=1, \ldots, p)$. A higher $\hat{w}_j$ indicates less penalization on the corresponding estimate. Clearly, the minimization in \eqref{est-reweighted-l1} preserves convexity with respect to $b$, and thus can be numerically solved as efficiently as the standard $l_1$ minimization. 

Equally importantly, by leveraging the posterior expectations $\hat{w}_j$, \bd\ provides a simple yet principled way to estimate the weights $w_j(\beta)$. The key implication from this is that the first-stage Bayesian modeling enables an informed second-stage estimation through its probabilistic characterization of the true parameter. 

Within the class of reweighted $l_1$ penalties, we introduce two specific instances illustrating how estimation and selection properties can be improved by reducing penalization on stronger signals.

\begin{enumerate}
  \item \textit{False-discovery (\fd) penalty}: Defined by 
  \begin{equation}\label{penalty-fd-l1}
    \mathcal{P}(\beta,b)=\sum_{j=1}^p\mathbbm{1}(\beta_j=0)|b_j|,
  \end{equation}
  the \fd\ penalty conveys the principle that penalization should ideally only target falsely included signals. Under the spike-and-slab prior to be introduced in \eqref{prior:ss}, it leads to an estimate with a well-interpretable form: 
  \begin{equation}\label{est-fd-l1}
    \est^\fd_\lambda=\mathop{\arg\min}_{b\in\mathbb{R}^p}\;\|X\bar{\beta}-Xb\|_2^2+\lambda\sum_{j=1}^p (1-\pi_j)|b_j|,
  \end{equation}
  where $\pi_j=pr(\gamma_j=1\mid y)$ is the marginal posterior inclusion probability (\textsc{pip}) for the $j$th predictor. Here, the weight estimates are $\hat{w}_j=1-\pi_j\ (j=1, \ldots, p)$, meaning that predictors with higher \textsc{pip}s receive lower shrinkage in the second-stage estimation. 

  \item \textit{Inverse-signal (\is) penalty}: Defined by 
  \begin{equation}\label{penalty-is-l1}
    \mathcal{P}(\beta,b)=\sum_{j=1}^p\frac{1}{|\beta_j|+\epsilon}|b_j|,\quad \epsilon\ge 0,
  \end{equation} 
  the \is\ penalty produces the estimate 
  \begin{equation}\label{est-is-l1}
    \est^\is_\lambda=\mathop{\arg\min}_{b\in\mathbb{R}^p}\;\|X\bar{\beta}-Xb\|_2^2+\lambda\sum_{j=1}^pE\left\{\frac{1}{|\beta_j|+\epsilon}\mid y\right\}|b_j|,
  \end{equation}
  where the weight estimates are $\hat{w}_j=E\{1/(|\beta_j|+\epsilon)\mid y\}\ (j=1, \ldots, p)$. This penalty is motivated by \cite{zou_adaptive_2006} and \cite{candes_enhancing_2008} who use weights inversely proportional to the true signal magnitude. Here $\epsilon$ is a small nonnegative constant introduced to guarantee the existence of the posterior expectation, and to bring in computational stability. 
\end{enumerate}

Note that the \fd\ penalty \eqref{penalty-fd-l1} and the \is\ penalty \eqref{penalty-is-l1} are suited to slightly different modeling scenarios. The penalty \eqref{penalty-fd-l1} has to be combined with selection priors which assign a non-zero probability mass at $\beta_j=0$. An example considered in this work is the classical spike-and-slab prior \citep{mitchell_bayesian_1988, george_variable_1993, george_approaches_1997, ishwaran_spike_2005} represented as 
\begin{equation}\label{prior:ss}
  \begin{aligned}
    \beta_\gamma\mid\gamma,\sigma^2&\sim N(0,\sigma^2n(X_\gamma^\top X_\gamma)^{-1}),\quad \beta_{1-\gamma}\mid\gamma=0\\
    \gamma_j\mid\pi_0&\sim \text{Bernoulli}(\pi_0)\quad  (j=1, \ldots, p),\\
    \pi_0&\sim \text{Beta}(1,1).
  \end{aligned}
\end{equation}
Here, $\gamma=[\gamma_1, \ldots, \gamma_p]^\top\in\{0,1\}^p$ is an inclusion indicator vector, where $\gamma_j=1$ (or 0) indicates the inclusion (or exclusion) of the $j$th predictor, i.e., $\beta_j\neq 0$ (or $\beta_j=0$). $X_\gamma$ is a matrix whose columns are the columns of $X$ corresponding to the included variables, with $\beta_\gamma$ being the corresponding coefficient vector. 

In contrast, the penalty \eqref{penalty-is-l1} applies well to both selection and shrinkage priors. An example of the latter is the horseshoe prior \citep{carvalho_handling_2009, carvalho_horseshoe_2010}, a popular member of the family of global-local shrinkage priors \citep{bernardo_shrink_2011}, represented as 
\begin{equation}\label{prior:hs}
  \beta_j\mid\eta_j,\tau,\sigma^2\sim N(0,\sigma^2\eta_j^2\tau^2),\quad\eta_j\sim C^+(0,1),\quad\tau\sim C^+(0,1),
\end{equation}
for $j=1, \ldots, p$, where $C^+(\cdot)$ denotes the half-Cauchy distribution. 

We note here that \cite{hahn_decoupling_2015} have explored a similar reweighted $l_1$ minimization, $\mathop{\min}_b\;\|X\bar{\beta}-Xb\|_2^2+\lambda\sum_{j=1}^p|b_j|/|\tilde{b}_j|$, with posterior mean $\bar{\beta}_j$ substituted for $\tilde{b}_j$. In our view, the formulation introduced in this work offers a more principled way of achieving the same reweighted penalization through a general class of loss functions. The formulation of \cite{hahn_decoupling_2015}, using direct substitution, is instead similar in spirit to the Iterative $l_1$ penalization method \citep{candes_enhancing_2008}, though \cite{hahn_decoupling_2015} use only a one-step method. The Iterative $l_1$ is considered as one of the competitors to \bd\ in our numerical experiments (Sections \ref{sec:result1} and \ref{sec:result2}).

\subsection{Statistical properties in the orthogonal design}\label{sec:property}

According to \cite{fan_li_2001}, a good penalty should result in an estimate that is sparse, continuous in the input data, and nearly unbiased when the true unknown signal is large. In this section, we establish relevant results for the \fd\ and \is\ penalties in the orthogonal design ($X^\top X=I_p$), or equivalently, in the normal means model $y_i\sim N(\beta_i,\sigma^2)\ (i=1, \ldots, n)$. To obtain an analytic form of the posterior, our discussion in this section is limited within the spike-and-slab prior \eqref{prior:ss} with fixed prior parameters $\sigma^2$ and $\pi_0$. In this case, as functions of $y_i$, the \textsc{pip} and posterior mean are 
$$
\pi_i=\xi_0/\left[\xi_0+(1+n)^{1/2}\exp\{-n_0y_i^2/(2\sigma^2)\}\right],\quad \bar{\beta}_i=n_0\pi_iy_i\quad(i=1, \ldots, n),
$$
where $\xi_0=\pi_0/(1-\pi_0)$, $n_0=n/(1+n)$. 

Under the orthogonal design, \bd\ with reweighted $l_1$ penalty applies what is commonly known as the soft-thresholding operation to the posterior mean $\bar{\beta}_i$, i.e., 
\begin{equation}\label{est-orthogonal}
\hat{\beta}_{\lambda,i}^{\text{Reweighted}}=\text{sgn}(\bar{\beta}_i)\left(|\bar{\beta}_i|-\lambda\hat{w}_i/2\right)_+.
\end{equation}
This achieves simultaneous sparsity and shrinkage: predictors with $|\bar{\beta}_i|<\lambda\hat{w}_i/2$ are excluded from the model, while predictors with $|\bar{\beta}_i|\ge\lambda\hat{w}_i/2$ have their coefficient estimates shrunk by $\lambda\hat{w}_i/2$ from the posterior mean $\bar{\beta}_i$ towards zero. Furthermore, the continuity of the estimate \eqref{est-orthogonal} with respect to the observed data $y_i$ can be established if the posterior quantities $\bar{\beta}_i$ and $\hat{w}_i$ are continuous in $y_i$. This condition is satisfied by the spike-and-slab prior under consideration when combined with the \fd\ penalty ($\hat{w}_i=1-\pi_i$) and the \is\ penalty ($\hat{w}_i=E\{1/(|\beta_i|+\epsilon)\mid y_i\}$). 

\begin{figure}
\figuresize{0.7}
\figurebox{20pc}{25pc}{}[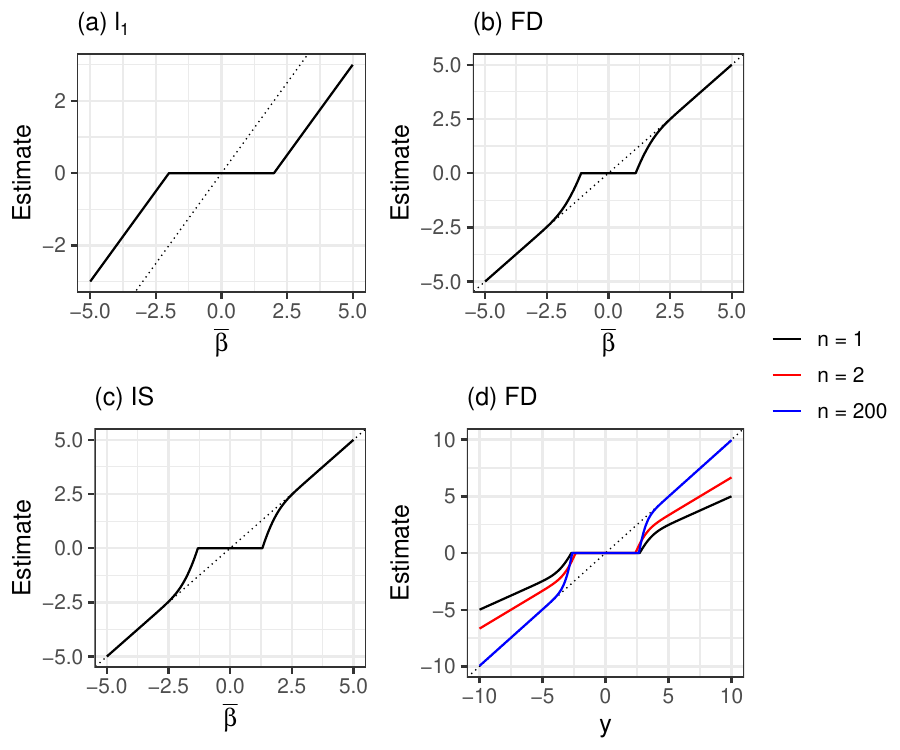]
\caption{Shrinkage function of \bd\ under the spike-and-slab prior with fixed prior parameters $\pi_0=0.5,\sigma^2=1$. The parameters used for illustration are $\lambda=2$ for the $l_1$ penalty, $\lambda=6$ for the \fd\ penalty, $\lambda=0.01$ and $\epsilon=0.001$ for the \is\ penalty. Each plot shows a 45-degree dotted reference line.}
\label{fig:shrinkage}
\end{figure}

\bd\ does not aim for exact unbiasedness, due to the nature of the first-stage Bayesian shrinkage. Yet, systematic bias arising from a poorly designed penalty in the second stage (e.g., the standard $l_1$ penalty) should be avoided. In this sense, the \fd\ and \is\ penalties debias the second stage by having $\hat{w}_i$ vanish to zero as the data $y_i$ increases in magnitude. The resulting estimate \eqref{est-orthogonal} approximately recovers the posterior mean $\bar{\beta}_i$ for large signals. If, in addition, the sample size $n$ is large enough, then \eqref{est-orthogonal} is nearly unbiased by producing $\hat{\beta}_{\lambda,i}^{\text{Reweighted}}\approx y_i$. These results are illustrated in Fig.~\ref{fig:shrinkage}.

\begin{figure}
\figuresize{0.5}
\figurebox{20pc}{25pc}{}[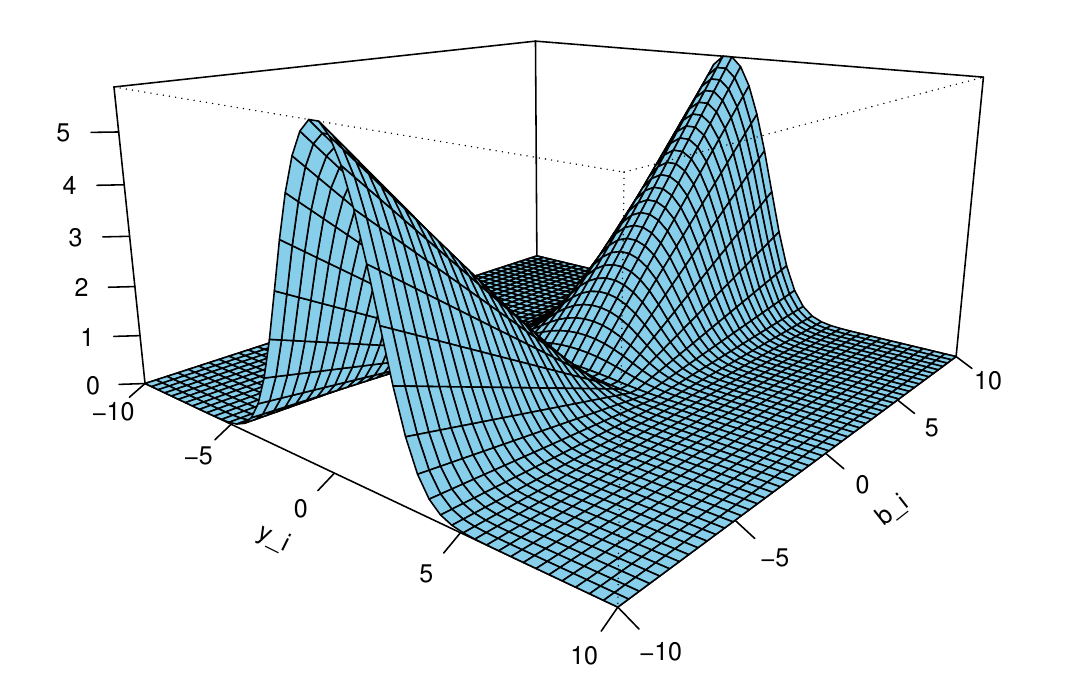]
\caption{Posterior expected \fd\ penalty function, under the spike-and-slab prior with parameters $\pi_0=0.5,\sigma^2=1,n=1$.}
\label{fig:penalty}
\end{figure}

As introduced in Section \ref{sec:penalty}, an important advantage of \bd\ is that debiasing can be achieved while preserving the convexity in the estimate $b$. This feat is possible because the penalty term, which usually depends only on the estimate $b$, is now expanded to a function of both $b$ and the unknown parameter $\beta$. Consequently, the posterior expected penalty establishes an automatic dependence on the observed data. Figure~\ref{fig:penalty} illustrates the posterior expected \fd\ penalty function in the orthogonal design (i.e., $(1-\pi_i)|b_i|$). It shows that the expected penalty decreases for large $|y_i|$ and is convex in $b_i$. The \is\ penalty behaves similarly, with numerical illustration omitted here.

The debiasing effect extends naturally to general non-orthogonal designs, as evidenced by the superior selection performance of the \fd\ and \is\ penalties compared to the standard $l_1$ penalty. This will be illustrated in our numerical results in Sections \ref{sec:result1} and \ref{sec:result2}.

\subsection{Tuning parameter selection by posterior benchmarking}\label{sec:post_benchmarking}

As our approach generates a path of estimates with varying degrees of sparsity, a question of importance is how to choose the most appropriate estimate along the solution path. Typically, an estimate is considered good if it demonstrates adequate predictive ability in relation to the true parameter, which is assumed to be sparse in our context. However, knowledge about the true parameter is always lacking. While denser estimates often achieve better predictive performance, they can introduce issues including overfitting and inclusion of irrelevant variables. Indeed, a central challenge of sparse estimation theory is balancing predictive ability with sparsity gain, especially in the absence of information about the true sparsity structure. 

In this regard, the Bayesian paradigm offers a unique advantage: the posterior mean provides a gold standard for prediction, based on which one can benchmark the loss in predictive power in pursuit of sparsity. Specifically, we define a metric to quantify squared prediction error of estimate $\est_\lambda$ as 
$$
  \mathcal{E}_\lambda=\mathcal{E}_\lambda(\beta,\est_\lambda)=\|X\beta-X\est_\lambda\|_2^2.
$$
The posterior mean $\bar{\beta}$ minimizes the posterior expected squared prediction error. We denote this benchmark as 
$$
\mathcal{E}=\inf_{b\in\mathbb{R}^p}E\{\|X\beta-Xb\|_2^2\mid y\}=E\{\|X\beta-X\bar{\beta}\|_2^2\mid y\}. 
$$
Increased sparsity leads to diminished predictive power compared to the dense estimate $\bar{\beta}$, as reflected in the posterior expectation of $\mathcal{E}_\lambda$. To determine an appropriate balance, we propose a \textit{posterior benchmarking criterion}: select the sparsest estimate for which the 90\% quantile-based credible interval of $\mathcal{E}_\lambda$ contains the benchmark $\mathcal{E}$. It means that the chosen sparse estimate should maintain predictive power closely aligned with the optimal posterior mean level. Using this criterion, the trade-off between prediction and sparsity is determined in an automatic, data-adaptive manner.

\begin{figure}
\figuresize{0.7}
\figurebox{20pc}{25pc}{}[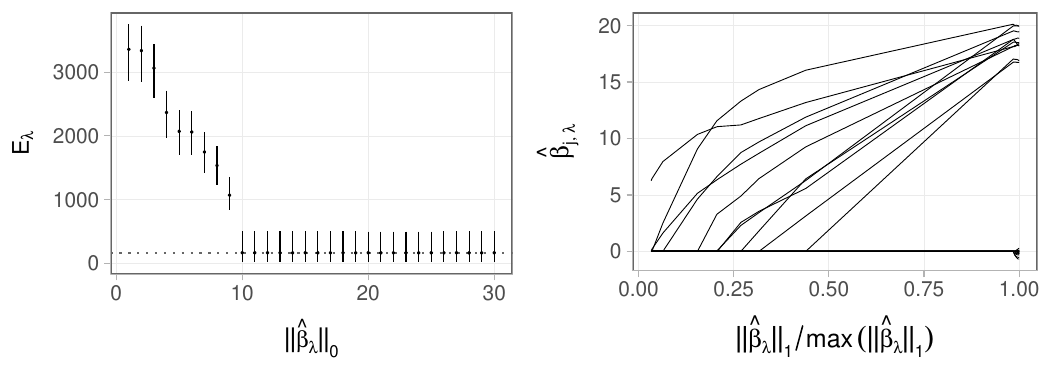]
\caption{Squared prediction error $\mathcal{E}_\lambda$ (left) and solution path (right) of \bd\ with \fd\ penalty. Displayed is one synthetic data under $n=50,p=30,s^*=20,k=10,\rho=0$. The plot of $\mathcal{E}_\lambda$ shows the posterior means of $\mathcal{E}_\lambda$ in dots and the 90\% credible intervals in bars; the horizontal dotted line represents the benchmark $\mathcal{E}$. The plot of $\hat{\beta}_{j,\lambda}$ shows different coefficients by different lines.}
\label{fig:lambda}
\end{figure}

As an example, Fig.~\ref{fig:lambda} (left) visualizes the path of $\mathcal{E}_\lambda$ against selected model size $\|\est_\lambda\|_0$ of \bd\ with \fd\ penalty on a synthetic dataset generated according to \eqref{simulation} with $n=50,p=30,s^*=20,k=10,\rho=0$. The plot demonstrates the general trend that denser \bd\ estimates have lower posterior expected squared prediction error. Additionally, the path exhibits a clear turning point at the true size $k=10$, where the posterior mean of $\mathcal{E}_\lambda$ approximates the benchmark $\mathcal{E}$. Beyond this point, further inclusion of predictors results in minimal reduction in $\mathcal{E}_\lambda$. In this instance, the sparse estimate selected by the posterior benchmarking criterion corresponds to a model size of $10$, revealing the true model size.

While our sparse estimation approach and regularized least squares (e.g., the lasso) share a similar form of the solution path (Fig.~\ref{fig:lambda} (right)), the dense solution of regularized least squares (e.g., taking $\lambda=0$) fails to offer a prediction benchmark comparable to the posterior mean. Consequently, methods like cross-validation are needed to determine the optimal estimate along the path. In such applications, however, there is no easy way to quantify how much predictive power is being lost due to sparsification relative to the optimal level.

A similar metric called variation-explained was proposed in \cite{hahn_decoupling_2015}. Both $\mathcal{E}_\lambda$ and variation-explained measure prediction accuracy. However, in our experience, the variation-explained rule is not always effective in high-dimensional problems.

\section{A reformulation of Bayesian variable selection}\label{sec:apm}

The classical Bayesian approach to learning a sparse model takes on the viewpoint of variable selection using spike-and-slab prior \eqref{prior:ss}. After deriving the posterior, the sparse estimate $\est$ is obtained by optimizing the associated model, $\hat{\gamma}\in\{0,1\}^p$, according to some decision-theoretic criterion among all candidate models. A prominent example of such an approach is the median probability model (\mpm) rule. It minimizes the posterior expected misclassification loss,
\begin{equation}\label{loss-mpm}
  L(\gamma, \hat{\gamma})=\sum_{j=1}^p\mathbbm{1}(\gamma_j\neq \hat{\gamma}_j)
\end{equation}
resulting in the model estimate $\hat{\gamma}^\mpm=[\mathbbm{1}(\pi_j\ge 1/2)]_{j=1}^p$. The corresponding sparse estimate of $\beta$ is taken to be the conditional posterior mean, denoted  $\est_{\hat{\gamma}^\mpm}=E\{\beta\mid y,\gamma=\hat{\gamma}^\mpm\}$. The \mpm\ rule is computationally easy and has been shown to achieve optimal prediction under the orthogonal design \citep{barbieri_optimal_2004}, making it a popular Bayesian approach to sparse learning. 

\begin{figure}
\figuresize{0.6}
\figurebox{20pc}{25pc}{}[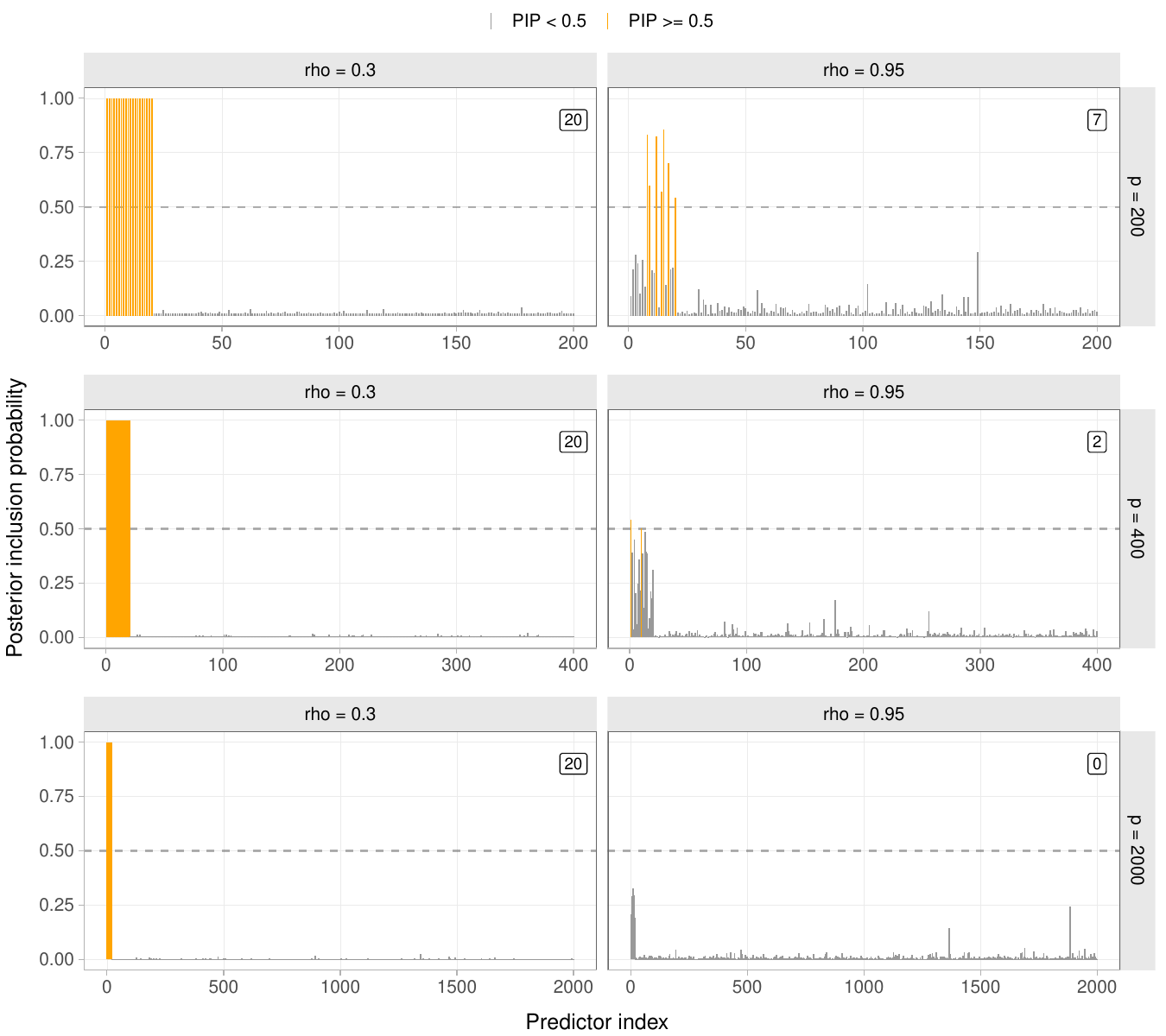]
\caption{Posterior inclusion probabilities across all predictors on synthetic data with $n=200$, $k=20$, $s^*=20$, $p=200,400,2000$, and $\rho=0.3,0.95$. Each plot is labeled by the number of predictors with $\textsc{pip}\ge 0.5$.}
\label{fig:pip}
\end{figure}

However, it has been observed that the performance of the \mpm\ rule deteriorates in high-dimensional, highly correlated settings. In such cases, the explanatory power is distributed across correlated variables, making the spike-and-slab posterior fail to concentrate within the model space. This is reflected in \textsc{pip}s being dispersed across variables, with very few exceeding the threshold $1/2$. Figure~\ref{fig:pip} illustrates such instances using synthetic data generated according to \eqref{simulation}. It includes an extreme case where all \textsc{pip}s fell below $1/2$ due to high correlation ($\rho=0.95$) and high dimensionality ($p=2000$), despite the true number of nonzero signals being $k=20$. Thus, chances are that the \mpm\ rule under-selects and suggests a much smaller model than needed \citep[see also][for similar observations]{ghosh_bayesian_2015}. Clearly, a remedy is to replace the fixed $1/2$ threshold on \textsc{pip}s with one that adapts to the posterior profile. Yet, determining the adaptive threshold introduces an additional challenge which remains unsolved in the literature.

The two-stage framework proposed in Section \ref{sec:sparse_est} suggests an immediate solution: the \mpm\ selection process can be integrated into our estimation scheme, so that it is able to produce a solution path of sparse models and allow for adaptive model selection aligned with the posterior profile. To achieve this, we advocate the following generalization of the symmetric 0-1 loss \eqref{loss-mpm}:
\begin{equation}\label{loss-apm}
  L_\lambda(\gamma, \hat{\gamma})=\sum_{j=1}^p\left[\lambda \mathbbm{1}(\gamma_j=0,\hat{\gamma}_j=1)+(1-\lambda)\mathbbm{1}(\gamma_j=1,\hat{\gamma}_j=0)\right],\quad\lambda\in [0,1].
\end{equation}
Here, the misclassification count is decomposed into false positives (the first term) and false negatives (the second term). Between the two error measures, a trade-off is allowed through a tuning parameter $\lambda$. Setting $\lambda$ closer to $0$ (or 1) lowers the cost of making false positives (or false negatives), thus encouraging a larger (or smaller) model. In this sense, the parameter $\lambda$ here parallels the role of $\lambda$ in the \bd\ framework \eqref{loss-bd}, enabling estimates with varying sparsity degrees to be generated. Clearly, when $\lambda=1/2$, the loss \eqref{loss-mpm} for the \mpm\ is recovered up to a constant factor as a special case. 

Importantly, the tuning parameter $\lambda$ in \eqref{loss-apm} directly corresponds to the threshold applied to posterior inclusion probabilities. To see this connection, derive the Bayes risk of \eqref{loss-apm} under the spike-and-slab prior \eqref{prior:ss} as
$$
E\{L_\lambda(\gamma, \hat{\gamma})\mid y\}=\sum_{j=1}^p\{\lambda\mathbbm{1}(\hat{\gamma}_j=1)(1-\pi_j)+(1-\lambda)\mathbbm{1}(\hat{\gamma}_j=0)\pi_j\}. 
$$
Due to its additive structure, it can be easily shown that the minimizer of the Bayes risk, $\hat{\gamma}_\lambda^\apm$, is coordinatewise
$$
  \hat{\gamma}_{\lambda,j}^\apm=\mathbbm{1}(\pi_j\ge \lambda)\quad (j=1, \ldots, p).
$$
Thus, the decision rule under the loss \eqref{loss-apm} boils down to a thresholding approach with the threshold determined by the tuning parameter $\lambda$: all and only variables with \textsc{pip}s greater than or equal to $\lambda$ are selected. We call $\hat{\gamma}_\lambda^\apm$ the \textit{adaptive probability model (\apm)}, emphasizing its contrast with \mpm\ which uses a fixed threshold $1/2$ on \textsc{pip}s.

Based on the selected model $\hat{\gamma}_\lambda^\apm$, the estimation of $\beta$ is completed by adopting the conditional posterior mean, analogous to the approach used in the \mpm\ rule, i.e., 
\begin{equation}\label{est-apm}
  \est_{\hat{\gamma}^\apm}=E\{\beta\mid y,\gamma=\hat{\gamma}^\apm\}.
\end{equation}
We call \eqref{est-apm} the \textit{\apm\ estimate} of $\beta$. This minimizes the posterior expected squared error loss $\|y^*-X^*b\|^2$ conditional on the selected model $\hat{\gamma}_\lambda^\apm$. 

Due to the equivalence between the tuning parameter $\lambda$ and the threshold on \textsc{pip}s, the problem of choosing the threshold translates directly to the problem of selecting $\lambda$, or equivalently, selecting an estimate from the solution path generated by varying $\lambda$. Furthermore, as this path is conceptually similar to the sparsification path produced by \bd, the selection can be addressed in a similar manner using the posterior benchmarking criterion introduced in Section \ref{sec:post_benchmarking}. As shown in Section \ref{sec:result1}, this approach can effectively address the issue of diminished selection power of \mpm\ arising from the reliance on a fixed threshold. 

\begin{figure}[ht]
  \figuresize{0.7}
  \figurebox{20pc}{25pc}{}[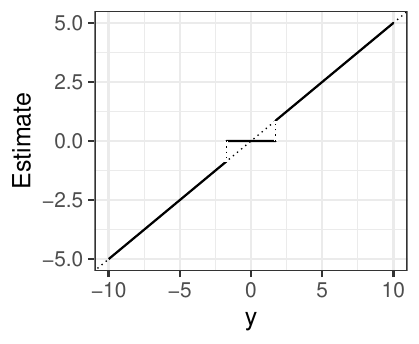]
  \caption{Shrinkage function of \apm\ under the spike-and-slab prior, with parameters $\pi_0=0.5,\sigma^2=1,n=1,\lambda=0.6$.}
  \label{fig:shrinkage-apm}
\end{figure}

A final remark on the difference between \apm\ and \bd\ is that \apm\ does not impose shrinkage on the (conditional) posterior mean of the selected variables, and is discontinuous in the input data, as shown in Fig.~\ref{fig:shrinkage-apm}. 

\section{Numerical experiments}\label{sec:result1}

Here we report results from a numerical experiment in which we assessed prediction and selection performance of \bd\ and \apm\ point estimates under various data generating settings. In each setting, we generated independent realizations of data $\{y,X\}$ according to
\begin{equation}\label{simulation}
  y\sim N(X\beta^*,I_n),\quad\beta^*=[s^*1_k^\top,0_{p-k}^\top]^\top,
\end{equation}
where $1_{k}$ is the vector of length $k$ filled with ones, $0_{p-k}$ is the zero vector of length $p-k$, and $s^*$ denotes the value of the nonzero signals. The rows of the design matrix $X=[x_1, \ldots, x_n]^\top$ were drawn independently as $x_i\sim N(0,R)\ (i=1,\ldots, n)$, with the $p \times p$ matrix $R$ having the $(i,j)$-th entry $n^{-1}\{\rho+(1-\rho)\mathbbm{1}(i=j)\}$ for some $\rho\ge 0$. In other words, all predictors were taken to share a common pairwise correlation $\rho \ge 0$. All realizations of $\{y,X\}$ were mean-centered before being put into any algorithm. 

In total, six different settings were considered, all with $n=200$, $k=20$ and $s^*=20$, but with one of three possible choices of $p \in \{200,400,2000\}$ -- representing moderate- to high-dimensional sparse settings, and one of two different choices of $\rho \in \{0.3,0.95\}$ -- corresponding to weak or strong predictor correlation. On such designs, it is known that the lasso fails to select the correct support with high probability \citep[see results in][for example]{sharp_wainwright}. Moreover, the high correlation renders the Bayesian \mpm\ method ineffective (as discussed in Section \ref{sec:apm}). We aim to investigate 1) whether the proposed approaches can recover the selection power of the \mpm\ in challenging settings, 2) how these methods compare to common regularized least squares methods such as the lasso, and 3) which combination of the considered priors and loss functions performs the best.

Posterior sampling was performed by using the R package \texttt{scaleBVS} \citep{zanella_scalable_2019} for the spike-and-slab prior \eqref{prior:ss} and the R package \texttt{horseshoe} \citep{bhattacharya_fast_2016} for the horseshoe prior \eqref{prior:hs}. Both implementations assume error variance prior $p(\sigma^2)\propto 1/\sigma^2$. For \bd, the optimization problems with $l_1$-type penalties described in Section \ref{sec:penalty} were solved by the R package \texttt{lars} \citep{efron_least_2004}. For \apm, the solution path was generated by leveraging its equivalence to a thresholding method on \textsc{pip}s, i.e., variables were selected in order of decreasing \textsc{pip}s. The tuning parameter $\lambda$ in both \bd\ and \apm\ was determined according to the posterior benchmarking criterion described in Section \ref{sec:post_benchmarking}. 

We evaluated an estimate $\est=[\hat{\beta}_j]_j$ using the following metrics: (a) selected model size, $\|\est\|_0$; (b) out-of-sample prediction mean squared error, $\mse=n^{-1}\|y^*-X^*\est\|_2^2$, where $\{y^*,X^*\}$ represents test data generated under the same design as $\{y,X\}$; (c) false discovery proportion, $\fdp=|\{j:\hat{\beta}_j\neq 0,\beta_j=0\}|/(|\{j:\hat{\beta}_j\neq 0\}|\vee 1)$, as a measure of Type I error; and (d) true positive proportion, $\tpp=|\{j:\hat{\beta}_j\neq 0,\beta_j\neq 0\}|/(|\{j:\beta_j\neq 0\}|\vee 1)$, as a measure of selection power.

Table \ref{table1} shows the performance of sparse estimates obtained by different methods across 500 independent replicates. Included for comparison are two regularized least squares methods related to our methods: the lasso and the Iterative $l_1$ algorithm \citep{candes_enhancing_2008}. The latter was implemented with $8$ iterations, using an adaptive choice of the stability number $\epsilon$, as described in Sections 2.2 and 3.2 of \cite{candes_enhancing_2008}. For both regularized least squares methods, the tuning parameter was chosen by minimizing 10-fold cross-validation \mse. 

\begin{table}[ht]
  \small
  \tbl{Selected model size, MSE, FDP and TPP averaged across 500 independent replicates with standard error in parentheses}
  {\begin{tabular}{@{}lrrrrrrrr@{}}
   & \multicolumn{4}{c}{$p=200, \rho=0.3$} & \multicolumn{4}{c}{$p=200, \rho=0.95$} \\ 
  & Size & MSE & FDP & TPP & Size & MSE & FDP & TPP \\ 
  MPM & 20.00{\tiny (0.00)} & 1.11{\tiny (0.12)} & 0.00{\tiny (0.00)} & 1.00{\tiny (0.00)} & 5.63{\tiny (1.38)} & 6.81{\tiny (2.93)} & 0.00{\tiny (0.02)} & 0.28{\tiny (0.07)} \\ 
  APM & 19.32{\tiny (0.47)} & 2.19{\tiny (0.80)} & 0.00{\tiny (0.00)} & 0.97{\tiny (0.02)} & 6.97{\tiny (1.20)} & 5.02{\tiny (1.05)} & 0.01{\tiny (0.04)} & 0.35{\tiny (0.06)} \\ 
  BD-$l_1$(SS) & 37.10{\tiny (5.33)} & 1.55{\tiny (0.48)} & 0.45{\tiny (0.08)} & 1.00{\tiny (0.00)} & 56.94{\tiny (6.14)} & 4.04{\tiny (0.65)} & 0.73{\tiny (0.04)} & 0.77{\tiny (0.10)} \\ 
  BD-FD(SS) & 20.00{\tiny (0.00)} & 1.11{\tiny (0.12)} & 0.00{\tiny (0.00)} & 1.00{\tiny (0.00)} & 9.09{\tiny (2.18)} & 5.33{\tiny (1.00)} & 0.03{\tiny (0.07)} & 0.44{\tiny (0.09)} \\ 
  BD-$l_1$(HS) & 47.68{\tiny (6.91)} & 1.22{\tiny (0.14)} & 0.57{\tiny (0.06)} & 1.00{\tiny (0.00)} & 61.97{\tiny (8.15)} & 1.47{\tiny (0.22)} & 0.67{\tiny (0.05)} & 0.99{\tiny (0.02)} \\ 
  BD-IS(HS) & 22.19{\tiny (3.51)} & 1.12{\tiny (0.12)} & 0.08{\tiny (0.11)} & 1.00{\tiny (0.00)} & 22.21{\tiny (3.53)} & 1.44{\tiny (0.22)} & 0.14{\tiny (0.11)} & 0.94{\tiny (0.05)} \\ 
  Lasso & 51.45{\tiny (5.30)} & 1.38{\tiny (0.17)} & 0.61{\tiny (0.04)} & 1.00{\tiny (0.00)} & 58.30{\tiny (6.35)} & 1.73{\tiny (0.27)} & 0.66{\tiny (0.04)} & 0.99{\tiny (0.02)} \\ 
  Iterative $l_1$ & 20.00{\tiny (0.00)} & 1.20{\tiny (0.14)} & 0.00{\tiny (0.00)} & 1.00{\tiny (0.00)} & 2.93{\tiny (0.75)} & 14.51{\tiny (3.96)} & 0.01{\tiny (0.05)} & 0.15{\tiny (0.04)} \\ 
  & \multicolumn{4}{c}{$p=400, \rho=0.3$} & \multicolumn{4}{c}{$p=400, \rho=0.95$} \\ 
  & Size & MSE & FDP & TPP & Size & MSE & FDP & TPP \\ 
  MPM & 20.00{\tiny (0.00)} & 1.11{\tiny (0.12)} & 0.00{\tiny (0.00)} & 1.00{\tiny (0.00)} & 3.80{\tiny (1.46)} & 19.02{\tiny (74.8)} & 0.01{\tiny (0.04)} & 0.19{\tiny (0.07)} \\ 
  APM & 19.31{\tiny (0.46)} & 2.21{\tiny (0.79)} & 0.00{\tiny (0.00)} & 0.97{\tiny (0.02)} & 7.52{\tiny (1.89)} & 4.87{\tiny (1.40)} & 0.04{\tiny (0.07)} & 0.36{\tiny (0.09)} \\ 
  BD-$l_1$(SS) & 46.34{\tiny (7.38)} & 1.62{\tiny (0.56)} & 0.56{\tiny (0.08)} & 1.00{\tiny (0.00)} & 70.94{\tiny (7.20)} & 4.66{\tiny (0.69)} & 0.82{\tiny (0.03)} & 0.65{\tiny (0.10)} \\ 
  BD-FD(SS) & 20.00{\tiny (0.00)} & 1.11{\tiny (0.12)} & 0.00{\tiny (0.00)} & 1.00{\tiny (0.00)} & 10.46{\tiny (3.88)} & 5.60{\tiny (1.18)} & 0.11{\tiny (0.14)} & 0.45{\tiny (0.11)} \\ 
  BD-$l_1$(HS) & 63.61{\tiny (12.4)} & 1.25{\tiny (0.15)} & 0.67{\tiny (0.06)} & 1.00{\tiny (0.00)} & 79.17{\tiny (9.96)} & 1.62{\tiny (0.29)} & 0.75{\tiny (0.04)} & 0.98{\tiny (0.04)} \\ 
  BD-IS(HS) & 27.75{\tiny (10.7)} & 1.15{\tiny (0.14)} & 0.21{\tiny (0.21)} & 1.00{\tiny (0.00)} & 28.29{\tiny (8.01)} & 1.59{\tiny (0.27)} & 0.30{\tiny (0.17)} & 0.93{\tiny (0.06)} \\ 
  Lasso & 64.26{\tiny (6.47)} & 1.53{\tiny (0.22)} & 0.69{\tiny (0.03)} & 1.00{\tiny (0.00)} & 74.74{\tiny (7.65)} & 2.06{\tiny (0.36)} & 0.74{\tiny (0.03)} & 0.97{\tiny (0.04)} \\ 
  Iterative $l_1$ & 20.00{\tiny (0.00)} & 1.21{\tiny (0.15)} & 0.00{\tiny (0.00)} & 1.00{\tiny (0.00)} & 2.67{\tiny (0.68)} & 16.00{\tiny (5.11)} & 0.02{\tiny (0.10)} & 0.13{\tiny (0.04)} \\ 
  & \multicolumn{4}{c}{$p=2000, \rho=0.3$} & \multicolumn{4}{c}{$p=2000, \rho=0.95$} \\ 
  & Size & MSE & FDP & TPP & Size & MSE & FDP & TPP \\ 
  MPM & 20.00{\tiny (0.00)} & 1.11{\tiny (0.12)} & 0.00{\tiny (0.00)} & 1.00{\tiny (0.00)} & 0.93{\tiny (0.94)} & 321.52{\tiny (362.9)} & 0.01{\tiny (0.08)} & 0.05{\tiny (0.05)} \\ 
  APM & 19.33{\tiny (0.47)} & 2.17{\tiny (0.80)} & 0.00{\tiny (0.00)} & 0.97{\tiny (0.02)} & 16.03{\tiny (5.61)} & 3.43{\tiny (1.24)} & 0.37{\tiny (0.18)} & 0.47{\tiny (0.12)} \\ 
  BD-$l_1$(SS) & 77.72{\tiny (11.7)} & 1.94{\tiny (0.92)} & 0.74{\tiny (0.05)} & 1.00{\tiny (0.00)} & 97.19{\tiny (7.37)} & 6.30{\tiny (0.88)} & 0.94{\tiny (0.02)} & 0.30{\tiny (0.10)} \\ 
  BD-FD(SS) & 20.00{\tiny (0.00)} & 1.11{\tiny (0.12)} & 0.00{\tiny (0.00)} & 1.00{\tiny (0.00)} & 34.76{\tiny (13.9)} & 6.05{\tiny (1.54)} & 0.72{\tiny (0.18)} & 0.40{\tiny (0.12)} \\ 
  BD-$l_1$(HS) & 96.50{\tiny (21.9)} & 1.44{\tiny (0.27)} & 0.78{\tiny (0.04)} & 1.00{\tiny (0.00)} & 119.47{\tiny (9.34)} & 2.97{\tiny (0.56)} & 0.89{\tiny (0.02)} & 0.66{\tiny (0.12)} \\ 
  BD-IS(HS) & 28.54{\tiny (25.1)} & 1.14{\tiny (0.17)} & 0.11{\tiny (0.23)} & 1.00{\tiny (0.00)} & 32.50{\tiny (15.0)} & 3.06{\tiny (0.69)} & 0.54{\tiny (0.24)} & 0.59{\tiny (0.13)} \\ 
  Lasso & 104.03{\tiny (9.88)} & 2.22{\tiny (0.48)} & 0.81{\tiny (0.02)} & 1.00{\tiny (0.00)} & 115.59{\tiny (10.5)} & 3.07{\tiny (0.53)} & 0.86{\tiny (0.03)} & 0.78{\tiny (0.11)} \\ 
  Iterative $l_1$ & 20.00{\tiny (0.00)} & 1.22{\tiny (0.16)} & 0.00{\tiny (0.00)} & 1.00{\tiny (0.00)} & 2.60{\tiny (0.65)} & 17.60{\tiny (5.15)} & 0.25{\tiny (0.31)} & 0.10{\tiny (0.04)} \\ 
  \end{tabular}}
  \label{table1}
  \begin{tabnote}
	MSE, mean square error; FDP, false discovery proportion; TPP, true positive proportion; BD-$l_1$, \bd\ with $l_1$ penalty; BD-FD, \bd\ with \fd\ penalty; BD-IS, \bd\ with \is\ penalty; SS, the spike-and-slab prior; HS, the horseshoe prior. 
  \end{tabnote}
\end{table}

Our first result is that, within the spike-and-slab framework, the selection power lost by \mpm\ in high-dimensional and highly correlated settings was effectively recovered by \bd\ or \apm.  As shown in Table \ref{table1}, when $\rho = 0.3$, \mpm\ achieved accurate support recovery with $\fdp=0$ and $\tpp=1$, yielding the lowest \mse\ among all methods. However, when $\rho=0.95$, the \tpp\ of \mpm\ decreased to 0.28 for $p=200$ and 0.05 for $p=2000$. In these cases, \bd\ and \apm\ significantly enhanced \tpp\ and reduced \mse, and such benefit increased as the dimension increased. 

Second, comparing the effects of different penalties within \bd\ under the same priors, we observe that \bd\ with $l_1$ penalty produced denser estimates with high \fdp, a performance similar to that of lasso. In contrast, \bd\ with \fd\ or \is\ penalties yielded sparser estimates with significantly lower \fdp. This result held for all settings. 

Third, we examine the three methods that similarly reweight the standard $l_1$ penalty by inverse signal magnitude. In weak correlation settings, these methods showed comparable predictive accuracy and selection power. A minor difference was that \bd\ with \is\ penalty under the horseshoe prior incurred more false discoveries. In strong correlation settings, however, clearer differences emerged. When $\rho=0.95$, \bd\ with \is\ penalty under the horseshoe prior achieved the highest selection power and lowest prediction error, without incurring a substantial increase in \fdp. \bd\ with \fd\ penalty under the spike-and-slab prior performed slightly worse in the sense that it showed higher \mse\ and a reduction in \tpp. Notably, the Iterative $l_1$ approach was ineffective in highly correlated settings; for all $p=200,400,2000$, it produced tiny models with high prediction error.

Finally, in general, \bd\ with \is\ penalty under the horseshoe prior had the best prediction and selection power among all methods compared. Across all settings, it attained lower \mse\ error than the lasso -- a method known for strong predictive performance but high false discovery rate -- while selecting noticeably fewer but more relevant predictors. On the other hand, if one prioritizes a lower false discovery proportion, then \apm\ would be the most preferable.

\section{Double regularization and improved solution path}\label{sec:result2}

\cite{su_false_2017} derived the asymptotic trade-off curve between \fdp\ and \tpp\ for the lasso, concluding that true features and null features are always interspersed on the lasso path in high-dimensional settings. The implication is remarkable: no estimate on such path can obtain ideal support recovery where all true positives are identified with no false discovery incurred. Motivated by this result, we performed an assessment of the proposed methods from the perspective of solution path improvement. In particular, we investigated how the implicit double regularization might improve the ordering of variables along the solution path compared to single-regularized paths such as those produced by the lasso, as well as those produced by more advanced iterative alternatives.

\begin{figure}[ht]
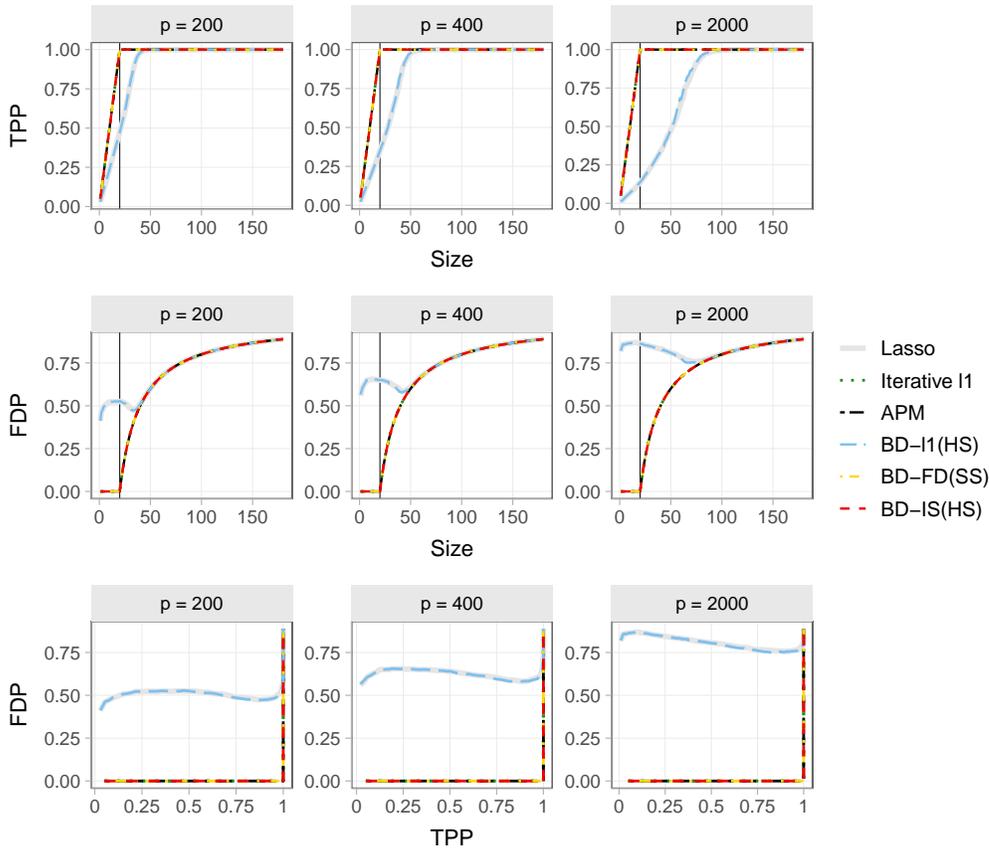

  \figuresize{0.7}
  \figurebox{20pc}{25pc}{}[Figures/Path\_rho3.pdf]
  \caption{Performance of the solution paths under $\rho=0.3$. The first and second rows show \tpp\ and \fdp\, respectively, averaged at different model sizes across 500 independent replicates, with vertical lines indicating the true model size $k = 20$. The third row shows the trade-off between \fdp\ and \tpp.}
  \label{fig:path_rho3}
\end{figure}

\begin{figure}[ht]
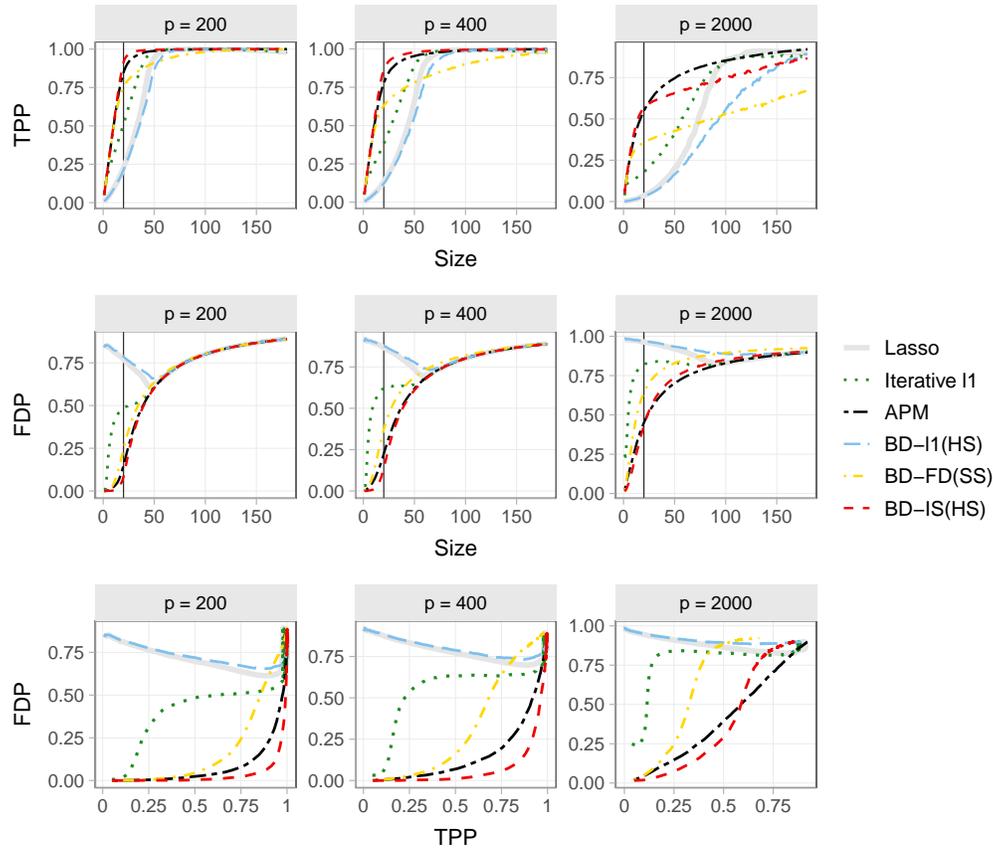

  \figuresize{0.7}
  \figurebox{20pc}{25pc}{}[Figures/Path\_rho95.pdf]
  \caption{Performance of the solution paths under $\rho=0.95$. The first and second rows show \tpp\ and \fdp\, respectively, averaged at different model sizes across 500 independent replicates, with vertical lines indicating the true model size $k = 20$. The third row shows the trade-off between \fdp\ and \tpp.}
  \label{fig:path_rho95}
\end{figure}

Figures \ref{fig:path_rho3} and \ref{fig:path_rho95} show results of a numerical comparison of solution paths from three \bd\ methods, \apm, lasso, and Iterative $l_1$, under the same settings as in Section \ref{sec:result1}, for $\rho = 0.3$ and $\rho = 0.95$ respectively. In each figure, the first and second rows show \tpp\ and \fdp, respectively, as model size increases along the path. The third row illustrates the trade-off between \fdp\ and \tpp\ to reflect the trade-off between Type I error and power. A good path should simultaneously achieve low Type I error and high power. For each of the methods compared, we computed and displayed solution paths covering model sizes from $1$ to $180$, rather than extending to the fully dense estimate to ease computation. For the Iterative $l_1$ algorithm, the solution path from the final iteration is shown. 

We observe that reweighting the standard $l_1$ penalty by inverse signal magnitude could substantially enhance the quality of the path by reducing the false discovery proportion early on the path. For example, in the case of $\rho=0.3,p=200$ shown in Fig.~\ref{fig:path_rho3}, \bd\ with $l_1$ penalty under the spike-and-slab prior had \fdp\ approximately $50\%$ before all true positives were picked up. This performance further deteriorated as the dimension $p$ grew. In contrast, for all $p=200,400,2000$, \bd\ with \fd\ penalty first selected all true signals without any false discoveries, achieving an optimal trade-off between \fdp\ and \tpp. The advantage of reweighting the $l_1$ penalty was also evident in the comparison between the lasso and the Iterative $l_1$ algorithm.

Among these peer methods, \apm\ under the spike-and-slab prior and \bd\ with \is\ penalty under the horseshoe prior offered the highest-quality solution paths, particularly in highly correlated designs. In particular, \bd\ with \is\ penalty under the horseshoe prior demonstrated strong selection power early on the path. As shown in Fig.~\ref{fig:path_rho95}, it achieved the highest \tpp\ within the first 20 selections, delivering the best \fdp-\tpp\ balance when $p = 200$ and $p = 400$. Yet, also note that this benefit weakened as the dimension increased to $p=2000$. In this case, \apm\ showed the quickest increase in \tpp\ after the first 20 selections, ultimately achieving the best balance with \fdp\ later on its path. 

Interestingly, the superior performance of \bd\ with \is\ penalty could not be replicated by the Iterative $l_1$ algorithm, despite their similar penalty constructions. While their performance was comparable in low-correlation settings ($\rho=0.3$), \bd\ with \is\ penalty demonstrated a clear advantage in high-correlation settings ($\rho=0.95$), achieving a significantly better \fdp-\tpp\ trade-off. Recall that both methods aim to debias the penalization through a first-step ordering of variable importance encoded by the inverse signal magnitude. However, the key difference lies in the estimation of the inverse signal weights: \bd\ uses the posterior expectation according to Bayesian decision theory, whereas the Iterative $l_1$ iteratively adopts the signal estimates from the previous iterations, initialized by a lasso solution. The much higher-quality solution paths of \bd\ implies the effectiveness of the former approach, in contrast to the latter which depends on less reliable estimates initialized by the lasso.

\section{Application: eQTL mapping data}\label{sec:eQTL}

This section presents a case study on data collected by \citet{Scheetz_2006} investigating gene regulation in the mammalian eye. They aimed to identify genetic variations relevant to human eye diseases using expression quantitative trait locus (eQTL) mapping in laboratory rats. The dataset contains expression values of over 31,000 gene probes measured in 120 male rat offsprings. One goal of the analysis was to identify genes that are associated with TRIM32 (probe ID 1389163\_at), a gene implicated in the human disease Bardet-Biedl syndrome. The dataset is publicly available at Gene Expression Omnibus (\texttt{www.ncbi.nlm.nih.gov/geo}; accession number GSE5680). 

\begin{figure}[ht]
  \figuresize{0.7}
  \figurebox{20pc}{25pc}{}[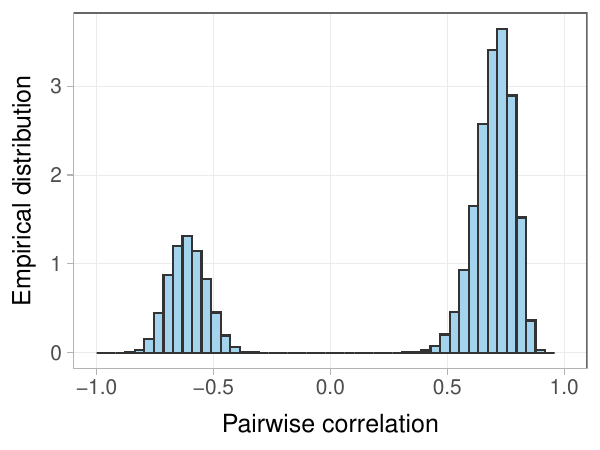]
  \caption{Distribution of pairwise correlation among 600 predictors in eQTL data.}
  \label{fig:eQTL_cor}
\end{figure}

Following routine preprocessing protocols for this dataset \citep[e.g.,][]{Scheetz_2006, Wang_quantile_2012, Peng_Wang_2015, huang2008adaptive}, we excluded probes that are not expressed or that lack sufficient variation, resulting in a subset of 18,976 probes, including the target probe of TRIM32. This subset was further narrowed to the top 600 probes with the highest absolute correlation with the expression values of TRIM32. Consequently, the data had sample size $n=120$ and dimension $p=600$. Moreover, among the 600 probes, typically only few are expected to be associated with TRIM32. Thus, this dataset presents a high-dimensional, correlated (as shown in Fig.~\ref{fig:eQTL_cor}), and sparse estimation problem. In this section, we apply and compare the proposed method as well as existing approaches to this problem.

To allow for performance assessment of the different methods, we first randomly split the 120 samples into a training set of size 100 and a test set of size 20. Expression values for all probes were standardized to have zero mean and unit $l_2$ norm. We implemented \mpm, \bd, \apm\ and regularized least squares methods as described in Section \ref{sec:result1}. Table \ref{table2} shows the selected model size and the prediction mean squared error for each method averaged across 500 random splits.

\begin{table}[ht]
  \tbl{Selected model size and MSE averaged across 500 random splits of the eQTL data with standard error in parentheses}
  {\begin{tabular}{@{}lrr@{}}
    & \multicolumn{1}{r}{Size} & \multicolumn{1}{r}{MSE} \\ 
    MPM & 1.266\ (1.091) & 0.039\ (0.014) \\ 
    APM & 2.738\ (1.485) & 0.031\ (0.014) \\ 
    BD-FD(SS) & 4.318\ (2.596) & 0.030\ (0.011) \\ 
    BD-IS(HS) & 4.396\ (3.602) & 0.032\ (0.013) \\ 
    Lasso & 42.686\ (13.23) & 0.024\ (0.011) \\ 
    Iterative $l_1$ & 17.120\ (6.265) & 0.026\ (0.012) \\ 
  \end{tabular}}
  \label{table2}
  \begin{tabnote}
    MSE, mean square error; BD-FD, \bd\ with \fd\ penalty; BD-IS, \bd\ with \is\ penalty; SS, the spike-and-slab prior; HS, the horseshoe prior. 
  \end{tabnote}
\end{table}

We observe that \bd\ and \apm\ produced significantly sparser estimates than the lasso and the Iterative $l_1$ algorithm. Specifically, among the Bayesian approaches compared, \mpm\ gave the smallest model with an average size of 1. As previously mentioned, this may be due to its reduced selection power on high-dimensional, correlated data. \bd\ with \fd\ and \is\ penalties yielded an average model size of 4 and showed better predictive power than \mpm. Lasso, by comparison, selected an average model size of 43. While the corresponding gain in predictive power might be beneficial, its necessity is debatable, as such denser estimates often include numerous small coefficient estimates that may be spurious from a variable selection standpoint.

\begin{figure}[ht]
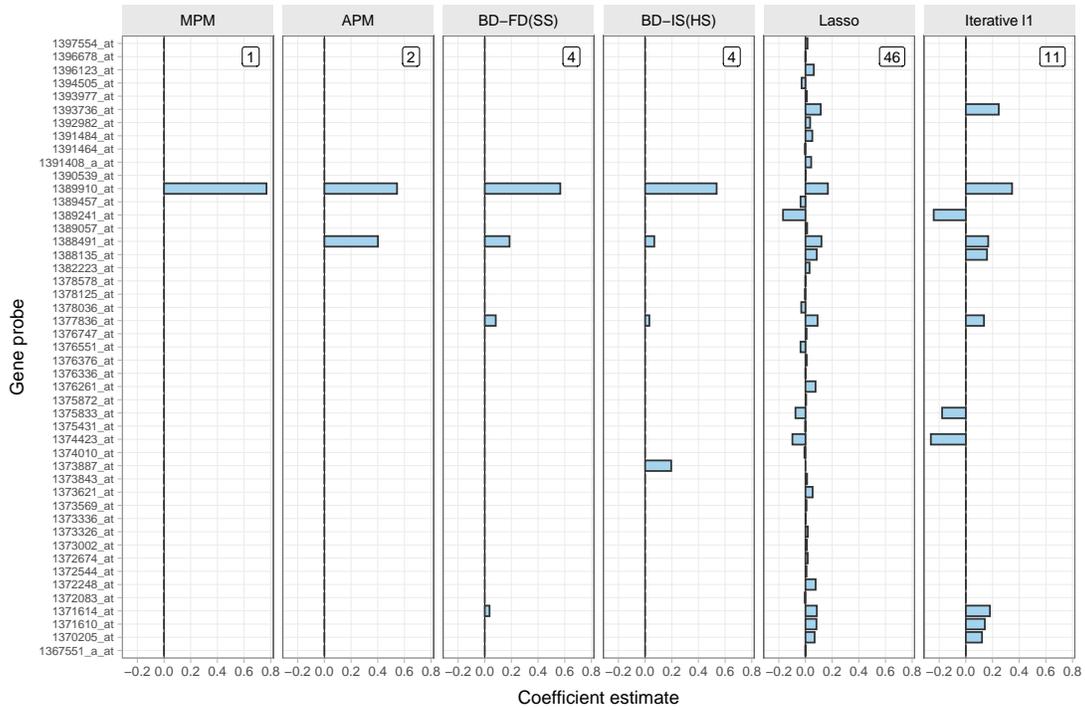

  \figuresize{0.63}
  \figurebox{20pc}{25pc}{}[Figures/eQTL\_est\_600.pdf]
  \caption{Coefficient estimates for probes selected by at least one of the compared methods, labeled by the selected model size.}
  \label{fig:eQTL_est}
\end{figure}

\begin{figure}[ht]
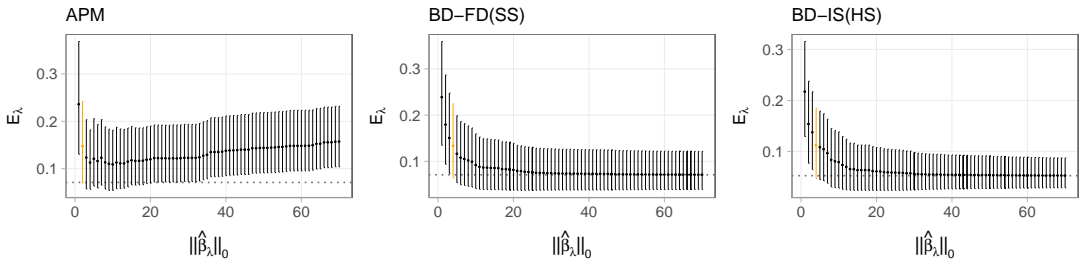

  \figuresize{0.57}
  \figurebox{20pc}{25pc}{}[Figures/eQTL\_dss\_600.pdf]
  \caption{Path of squared prediction error $\mathcal{E}_\lambda$ of \bd\ and \apm\ on the eQTL data, with the selected model size highlighted in orange.}
  \label{fig:eQTL_dss}
\end{figure}

Second, to identify key gene probes associated with TRIM32, we applied the methods to the full dataset with a sample size of 120. Figure~\ref{fig:eQTL_est} presents the coefficient estimates for probes selected by at least one of the compared methods. Observations regarding the selected model size were consistent with those in Table \ref{table2} which analyzed randomly split data. Figure~\ref{fig:eQTL_dss} displays the path of the squared prediction error, $\mathcal{E}_\lambda$, of \apm\ and \bd\ approaches. In both plots of \bd, the posterior mean of $\mathcal{E}_\lambda$ dropped rapidly with the first few selected predictors; however, this slowed down as the model became denser -- a typical behavior in high-dimensional, correlated data where predictors share common explanatory power. In the plot of \apm, the posterior mean of $\mathcal{E}_\lambda$ reached lower values by the first few selected predictors, while increasing again toward the denser end of the solution path. This behavior is due to the lack of proper shrinkage on the conditional posterior mean, as illustrated in Fig.~\ref{fig:shrinkage}. 

Based on the estimation results of \apm, \bd\ with \fd\ penalty under the spike-and-slab prior, and \bd\ with \is\ penalty under the horseshoe prior, we conclude the potential importance of the following probes: Probes 1389910\_at and 1388491\_at were selected by all three approaches, suggesting a significant association with TRIM32 expression; probe 1389910\_at, in particular, had the largest absolute coefficient estimate. Additionally, probe 1377836\_at was picked up by both \bd\ approaches, albeit with smaller estimates.  

\section{Discussion}\label{sec:discussion}

Bayesian Decoupling introduces a mathematically neat and computationally easy integration of Bayesian shrinkage with regularized least squares, contributing to advancements in both fields. From the Bayesian perspective, the use of a family of loss functions moves beyond the conventional reliance on a single symmetric loss, producing a solution path instead of a single sparse estimate. This framework offers an adaptive thresholding alternative to the current Bayesian variable selection method which has been criticized for using a fixed threshold that does not adapt well to diverse posterior scenarios. From the regularized estimation perspective, the probabilistic characterization of true parameters provided by Bayesian modeling brings significant advantages to penalty construction. In our reweighted penalization scheme, the simultaneous bias reduction and convexity, along with a decision-theoretic principle to estimate the weights, are benefits unparalleled by existing regularization methods developed outside of the Bayesian context. Also enabled by the Bayesian paradigm is the posterior benchmarking criterion to choose the tuning parameter; it replaces the cross-validation which is computationally expensive. Furthermore, our numerical experiments show very promising results. With an appropriately chosen combination of prior and loss, the doubly-regularized estimation produces high-quality solution paths that are not easily achievable by comparable regularized least squares approaches. The resulting sparse estimates show increased selection power, reduced false discoveries, and decreased prediction error compared to peer methods in a variety of high-dimensional settings with correlated predictors. 

Admittedly, this is not the first work that incorporates Bayesian elements into penalization techniques or vice versa \citep{Rockova_SSL,rovckova2016bayesian, Bondell_credible}. However, unique to our work is not only a novel formal synthesis but also a fundamentally new insight into regularized estimation theory. The unification of Bayesian shrinkage and penalization -- two regularization methods traditionally developed in parallel -- deserves consideration even beyond the sparse estimation context. In general situations with signals blurred by noisy measurements, Bayesian shrinkage can be used for data denoising and signal amplification, followed by the application of penalized loss to distill a desired regularized estimate.

Meanwhile, as an exploration of this novel approach, it raises more questions than it answers. Our current arguments are primarily empirical, and further theoretical development is needed. For example, an important direction is to examine the convergence properties of our approach relative to the minimax convergence rate established by \cite{donoho_nearly_black}, which is a widely recognized standard for assessing sparse estimators in both frequentist \citep[e.g.,][]{Abramovich_adapting, su_slope_2016} and Bayesian \citep[e.g.,][]{van_der_Pas_horseshoe} paradigms. Of particular interest is the utility of the proposed posterior benchmarking criterion for selecting $\lambda$ in achieving this rate. This is especially significant as attaining the minimax rate typically requires tuning parameters that encode knowledge about the true sparsity structure, while our method automates this process. 

We conclude the discussion with another extension of our approach incorporating false discovery rate (\textsc{fdr}) control, a growing focus in high-dimensional multiple testing problems \citep{bh_1995, storey2011false}. Specifically, due to the practical infeasibility of exact \textsc{fdr} computation in the frequentist sense, we consider its Bayesian variant, the \textit{local false discovery rate} \citep{efron_empirical_2001, efron_empirical_2002, storey_2003}, expressed as 
$$
\text{fdr}(\est_\lambda)=E\left\{\frac{|\{j:\hat{\beta}_{j,\lambda}\neq 0,\beta_j=0\}|}{|\{j:\hat{\beta}_{j,\lambda}\neq 0\}|\vee 1}\mid y\right\}=\frac{\sum_{j=1}^p(1-\pi_j)\mathbbm{1}(\hat{\beta}_{j,\lambda}\neq 0)}{|\{j:\hat{\beta}_{j,\lambda}\neq 0\}|\vee 1}
$$
under the spike-and-slab prior \eqref{prior:ss}. A denser estimate generally incurs more false discoveries. Accordingly, to control \textsc{fdr}, one could select the densest estimate for which the local false discovery rate remains below a prespecified threshold (e.g., 10\%). Given the established connection between local false discovery rate and frequentist's \textsc{fdr} in simple scenarios \citep[see e.g.,][]{efron_size_2007}, it would be interesting to explore whether this method effectively control \textsc{fdr} in general settings.

\bibliographystyle{biometrika}
\bibliography{main.bib}

\end{document}